\documentclass[letterpaper]{article} 
\usepackage{aaai25}  
\usepackage{times}  
\usepackage{helvet}  
\usepackage{courier}  
\usepackage[hyphens]{url}  
\usepackage{graphicx} 
\usepackage{natbib}  
\usepackage{caption} 
\usepackage{algorithm}
\usepackage{algorithmic}

\usepackage{amssymb}
\usepackage{amsmath}
\usepackage{subfigure}
\usepackage{multirow}

\usepackage{newfloat}
\usepackage{listings}
\DeclareCaptionStyle{ruled}{labelfont=normalfont,labelsep=colon,strut=off} 
\floatstyle{ruled}
\newfloat{listing}{tb}{lst}{}
\floatname{listing}{Listing}
\title{Personalized Federated Collaborative Filtering: \\A Variational AutoEncoder Approach}
\author{
    Zhiwei Li\textsuperscript{\rm 1}, 
    Guodong Long\textsuperscript{\rm 1}, 
    Tianyi Zhou\textsuperscript{\rm 2}, 
    Jing Jiang\textsuperscript{\rm 1}, 
    Chengqi Zhang\textsuperscript{\rm 3}    
}
\affiliations{
    \textsuperscript{\rm 1}Australian Artificial Intelligence Institute, Faculty of Engineering and IT, University of Technology Sydney\\
    \textsuperscript{\rm 2}Department of Computer Science, University of Maryland, College Park\\
    \textsuperscript{\rm 3}Department of Data Science and Artificial Intelligence, The Hong Kong Polytechnic University\\

    zhiwei.li@student.uts.edu.au,
    \{guodong.long, jing.jiang\}@uts.edu.au, 
    zhou@umiacs.umd.edu,
    chengqi.zhang@polyu.edu.hk

}

\begin{document}

\newtheorem{lemma}{Lemma}

\maketitle

\begin{abstract}
Federated Collaborative Filtering (FedCF) is an emerging field focused on developing a new recommendation framework with preserving privacy in a federated setting.
Existing FedCF methods typically combine distributed Collaborative Filtering (CF) algorithms with privacy-preserving mechanisms, and then preserve personalized information into a user embedding vector. 
However, the user embedding is usually insufficient to preserve the rich information of the fine-grained personalization across heterogeneous clients.
This paper proposes a novel personalized FedCF method by preserving users' personalized information into a latent variable and a neural model simultaneously.
Specifically, we decompose the modeling of user knowledge into two encoders, each designed to capture shared knowledge and personalized knowledge separately. A personalized gating network is then applied to balance personalization and generalization between the global and local encoders.
Moreover, to effectively train the proposed framework, we model the CF problem as a specialized Variational AutoEncoder (VAE) task by integrating user interaction vector reconstruction with missing value prediction. 
The decoder is trained to reconstruct the implicit feedback from items the user has interacted with, while also predicting items the user might be interested in but has not yet interacted with.
Experimental results on benchmark datasets demonstrate that the proposed method outperforms other baseline methods, showcasing superior performance.
Our code is available at https://github.com/mtics/FedDAE.

\end{abstract}

%

\section{Introduction}

In the digital age, Recommendation Systems have become essential tools for filtering online information and helping users discover products, content, and services that match their preferences \cite{ko2022survey}. 
Collaborative Filtering (CF) is widely recognized for its ability to generate personalized recommendations by analyzing the relationships between users and items based on user interaction data \cite{shen2020collaborative}. 
However, with the enforcement of data privacy laws like GDPR \cite{voigt2017eu}, safeguarding privacy has become increasingly critical.Traditional CF methods typically require centralizing user data on servers for processing, a practice that is no longer viable in today's privacy-conscious environment. 

To address this challenge, Federated Collaborative Filtering (FedCF) has emerged, combining the principles of federated learning (FL) and CF \cite{yang2020federated}. FedCF enables models to be trained on users' devices, eliminating the need to upload private data to central servers, thus ensuring data privacy while still providing recommendation services \cite{ammad2019federated}.
Existing work on FedCF \cite{chai2020secure,lin2020meta,lin2020fedrec} is mostly based on matrix factorization (MF) \cite{koren2009matrix}, which is favored for its computational efficiency and strong interpretability. 
MF effectively captures the latent relationships between users and items, making it widely used in many practical applications and delivering strong performance in recommendation systems \cite{mehta2017review}. 

However, the fundamental assumption of MF is that the relationships between users and items are linear, which may limit the model's ability to handle complex, non-linear relationships. The success of Neural Collaborative Filtering (NCF) \cite{he2017neural} demonstrates the necessarity of building non-linear relationships among users and items. In addition to FedNCF \cite{perifanis2022federated} which is a federated NCF algorithm, some recent works proposed to preserve the personalization on neural models in federated settings. 
PFedRec \cite{zhang2023dual} proposed using personalized item embeddings to capture each user's unique perspective about the relationships among items. 
FedRAP \cite{li2024federated} decomposes the item embedding to two additive components that preserve shared and personalized knowledge respectively. 
However, these methods rely on personalized item embedding that may lead to poor performance on model generalization, and maintaining a personalized item embedding is computation consuming.

\textbf{Main Contribution.} 
To effectively preserve fine-grained personalization and non-linear user-item relationships, we propose a novel \textbf{gating dual-encoder} structure that transforms the user-item interaction vector into two separate latent subspaces. 
Specifically, through collaborative training across clients, the global encoder maps the user profile into a universal latent subspace shared among clients. 
Meanwhile, the local encoder remains on the client side, where it is trained to map the user profile into a user-specific latent subspace. Consequently, the global encoder retains shared knowledge across clients, while the local encoder preserves personalized knowledge.
It is worth noting our proposed framework preserves personalized information at three levels. 
First, the user profile includes user-specific interaction records, allowing the global encoder to transform it into a user-specific vector within a universal latent subspace. 
Second, the personalized local encoder further transforms the user profile into a unique latent subspace. 
Third, the gating network adaptively adjusts the importance weights of the two encoders, balancing the performance of our proposed model between personalization and generalization.

To effectively train the proposed gating dual encoders, we revisit the FedCF problem from the perspective of Variational Autoencoders (VAEs) \cite{kingma2013auto} and propose a novel personalized FedCF method, which incorporates a gating dual-encoder VAE, named \textbf{FedDAE}.
To enhance the model's generalization capability, FedDAE employs a gating network that generates weights based on user interaction data, dynamically combining the outputs of the two encoders to achieve additive personalization. 
To train our method FedDAE, a global decoder is attached to the dual encoders, and the objective function of FedDAE is optimized through an designed alternating update process.

The paper's main contributions are summarized as below:
\begin{itemize}
    \item This work reformulates the FedCF problem as a VAE task to capture complex nonlinear relationships and maintain fine-grained personalization;
    \item We adopt a dual-encoder structure to separate shared and personalized knowledge, coupled with a personalized gating network that dynamically adjusts encoder weights, effectively balancing generalization and personalization to enhance FedCF tasks;
    \item Comprehensive experimental analyses are conducted to validate the effectiveness of FedDAE.
\end{itemize}

\section{Related Work}

\paragraph{Personalized Federated Learning.}
Standard federated learning methods, such as FedAvg \cite{mcmahan2017communication}, learn a global model on the server while considering data locality on each device \cite{li2024navigating}. However, these methods are limited in their effectiveness when dealing with non-IID data. PFL aims to learn personalized models for each device to address this issue \citep{tan2022towards}, often necessitating server-based parameter aggregation \cite{arivazhagan2019federated,t2020personalized,collins2021exploiting,li2024personalized,zhang2024beyond}. 
Several studies \cite{ammad2019federated,chen2023federated,chen2024personalized,yang2024dual} accomplish PFL by introducing various regularization terms between local and global models. 
Meanwhile, some work focus on personalized model learning by promoting the closeness of local models via variance metrics \cite{flanagan2020federated}, or enhancing this by clustering users into groups and selecting representative users for training \cite{li2021ditto,luo2022personalized,tan2023federated,yan2023personalization}, instead of random selection.
APFL \cite{deng2020adaptive} proposes adaptive PFL and derives the generalization bound for the mixture of local and global models, thereby achieving a balance between global collaboration and local personalization.

\paragraph{Federated Collaborative Filtering.}
As privacy protection becomes increasingly important, many studies \cite{hegedHus2019decentralized,chai2020secure,zhang2021vertical,zhang2024federated,zhang2024privfr,zhang2024transfr} have focused on FedCF. In this paper, we mainly focus on model-based personalized FedCF, which leverages users' historical behavior data, such as ratings, clicks, and purchases, to learn latent factors that represent user and item characteristics \cite{aggarwal2016model}.
To mitigate the impacts caused by client heterogeneity, personalized FedCF has received considerable attention due to its ability to take into account personalized information for each user. 
Early works \cite{lin2020fedrec,liang2021fedrec++,zhu2022cali3f,luo2023perfedrec++,yuan2023hetefedrec,zhang2023lightfr} primarily focused on modeling user preferences. In contrast, PFedRec \cite{zhang2023dual} and FedRAP \cite{li2024federated} have implemented dual personalization, which means they personalized both user preferences and item information. 
However, the nature of most existing work on matrix factorization limits their ability to model complex nonlinear relationships in data.

\paragraph{Variational Autoencoders.} 
VAEs have become increasingly important in recommendation systems due to their ability to model complex data distributions and handle data sparsity issues \cite{liang2024survey}. 
VAEs learn latent representations of user-item interaction data by mapping high-dimensional data to a low-dimensional latent space, which effectively captures complex patterns in user preferences and item characteristics \cite{kingma2013auto}. This latent space enables efficient encoding of both user and item information, facilitating accurate predictions of future interactions \cite{shenbin2020recvae}. 
Recent research has focused on enhancing VAE models to improve recommendation performance, including combining VAEs with Generative Adversarial Networks \cite{lee2017augmented,yu2019vaegan,gao2020zero}, integrating VAEs with more effective priors \cite{tomczak2018vae,klushyn2019learning,shenbin2020recvae,tran2024learning}, and enhancing VAEs with contrastive learning techniques \cite{aneja2021contrastive,xie2021adversarial,wang2022contrastvae}. 
These strategies have significantly boosted the performance of VAEs in recommendation tasks.
HI-VAE \cite{nazabal2020handling} has demonstrated the effectiveness of VAE in density estimation and missing data imputation through experiments on heterogeneous data completion tasks.
Mult-VAE \cite{liang2018variational} proves that using multinomial distribution as the likelihood function is more suitable for implicit feedback.
FedVAE \cite{polato2021federated} represents the first attempt to extend VAE to a federated setting; however, it aggregates gradients for all parameters on the server, which can lead to the loss of personalized information from individual clients.

\section{Problem Formulation}
Given $n$ users and $m$ items, let $\mathcal{U}$ and $\mathcal{I}$ be the sets of users and items respectively. 
We assume of knowing the users' implicit feedback, i.e., $\mathbf{R} = [\mathbf{r}_1, \mathbf{r}_2, \cdots, \mathbf{r}_n]^T \in \{0, 1\}^{n \times m}$,  where $r_{ui}=1$ if the $u$-th user interacted with the $i$-th item, and $r_{ui}=0$ otherwise. 
For the $u$-th user, we use $\mathbf{r}_u \in \{0, 1\}^{m}$ to indicate the interaction history, $\mathcal{I}_u$ to denote the set of all interacted items, and its length is $m_u$.

\subsection{Federated Collaborative Filtering (FedCF)}
In recommendation tasks with implicit feedback, Collaborative Filtering (CF) methods typically rely solely on users' interaction patterns \( \mathbf{R} \) to capture the relationships between users $\mathcal{U}$ and items $\mathcal{I}$, allowing for generating item recommendations without the need for exogenous information about the users or items \cite{koren2021advances}.
However, in the context of FL, the task of FedCF is to model the latent relationships between each user \( u \) and items $\mathcal{I}$ at their client using the observed portions of their interaction data \(\mathbf{r}_u\) locally. 
The goal is to generate predicted ratings \(\hat{\mathbf{r}}_u\) while simultaneously ensuring the protection of user privacy. Thus, we have the following for recommendation:
\begin{equation}
    \min \sum_{u=1}^n\mathcal{L}_{recon}(\hat{\mathbf{r}}_u, \mathbf{r}_u),
    \label{eq:recon}
\end{equation}
where \(\mathcal{L}_{recon}\) measures the reconstruction loss between \(\hat{\mathbf{r}}_u\) and \(\mathbf{r}_u\).
By optimizing Eq. \ref{eq:recon}, we ensure that \(\hat{\mathbf{r}}_u\) follows the same distribution as \(\mathbf{r}_u\), i.e., \(\hat{\mathbf{r}}_u \sim p(\mathbf{r}_u)\) \((u = 1,2,\dots,n)\).

\subsection{A VAE perspective for FedCF.}
From the perspective of VAEs, we first sample a latent variable \(\mathbf{z}_{u} \sim \mathcal{N}(0, \mathbf{I}_{k})\), which is assumed to model the $u$-th user's decision logic of the preference in recommendation system. 
It is worth noting that the work \cite{liang2018variational} suggested that multi-nominal distribution might be an appropriate option for recommendation with implicit feedback.

Given the sampled variable \(\mathbf{z}_{u}\), 
the user's preference probabilities for items can be generated by a nonlinear function \(f_{\mathbf{\theta}}: \mathbb{R}^{k} \rightarrow \mathbb{R}^m\), parameterized by \(\mathbf{\theta}\), applied to \(\mathbf{z}_{u}\):
\begin{equation}
    \pi(\mathbf{z}_{u}) = f_{\theta}(\mathbf{z}_{u}),
\end{equation}
s.t. $\mathbf{r}_u$ is drawn from $\mathbf{r}_{u} \sim \operatorname{Mult} (m_u, \pi (\mathbf{z}_{u}))$.
To measure the correctness of the predicted user preference, we use log-likelihood of $\mathbf{r}_u$ conditioned on $\mathbf{z}_{u}$ defined as follows:
\begin{equation}
    \log p_{\mathbf{\theta}} (\mathbf{r}_{u} | \mathbf{z}_{u})=\sum_{i=1}^{m} \mathbf{r}_{u i} \log \pi_{i}(\mathbf{z}_{u}).
    \label{eq:condition_likelihood}
\end{equation}

We apply variational inference \cite{jordan1999introduction} to approximate the intractable posterior distribution \(p(\mathbf{z}_{u} | \mathbf{r}_{u})\) using variational distribution to estimate \(\mathbf{\theta}\) for the function \(f_{\theta}\):
\begin{equation}
    q_{\mathbf{\Phi}_u} (\mathbf{z}_{u}| \mathbf{r}_{u})=
    \mathcal{N} (\mu_{\mathbf{\Phi}_u}(\mathbf{r}_{u}), \operatorname{diag}\{\sigma_{\mathbf{\Phi}_u}^{2}(\mathbf{r}_{u})\}).
    \label{eq:combine_posterior}
\end{equation}
Given \(k << \{n, m\}\), both of the vectors \(\mu_{\mathbf{\Phi}_u}(\mathbf{r}_{u}) \) and \(\sigma_{\mathbf{\Phi}_u}^{2}(\mathbf{r}_{u})\) in Eq. \eqref{eq:combine_posterior} are \(k\)-dimensional outputs of the data-dependent function \(g_{\mathbf{\Phi}_u}(\mathbf{r}_u) = [\mathbf{\mu}_{\mathbf{\Phi}_u}(\mathbf{r}_u), \mathbf{\sigma}_{\mathbf{\Phi}_u}(\mathbf{r}_u)] \in \mathbb{R}^{2k}\), where the parameter \(\mathbf{\Phi}_u\) is used to capture the representation of item features on the  \(u\)-th client \((u = 1,2,\dots,n)\).



\begin{figure}[!t]
    \centering
    \includegraphics[width=1\linewidth]{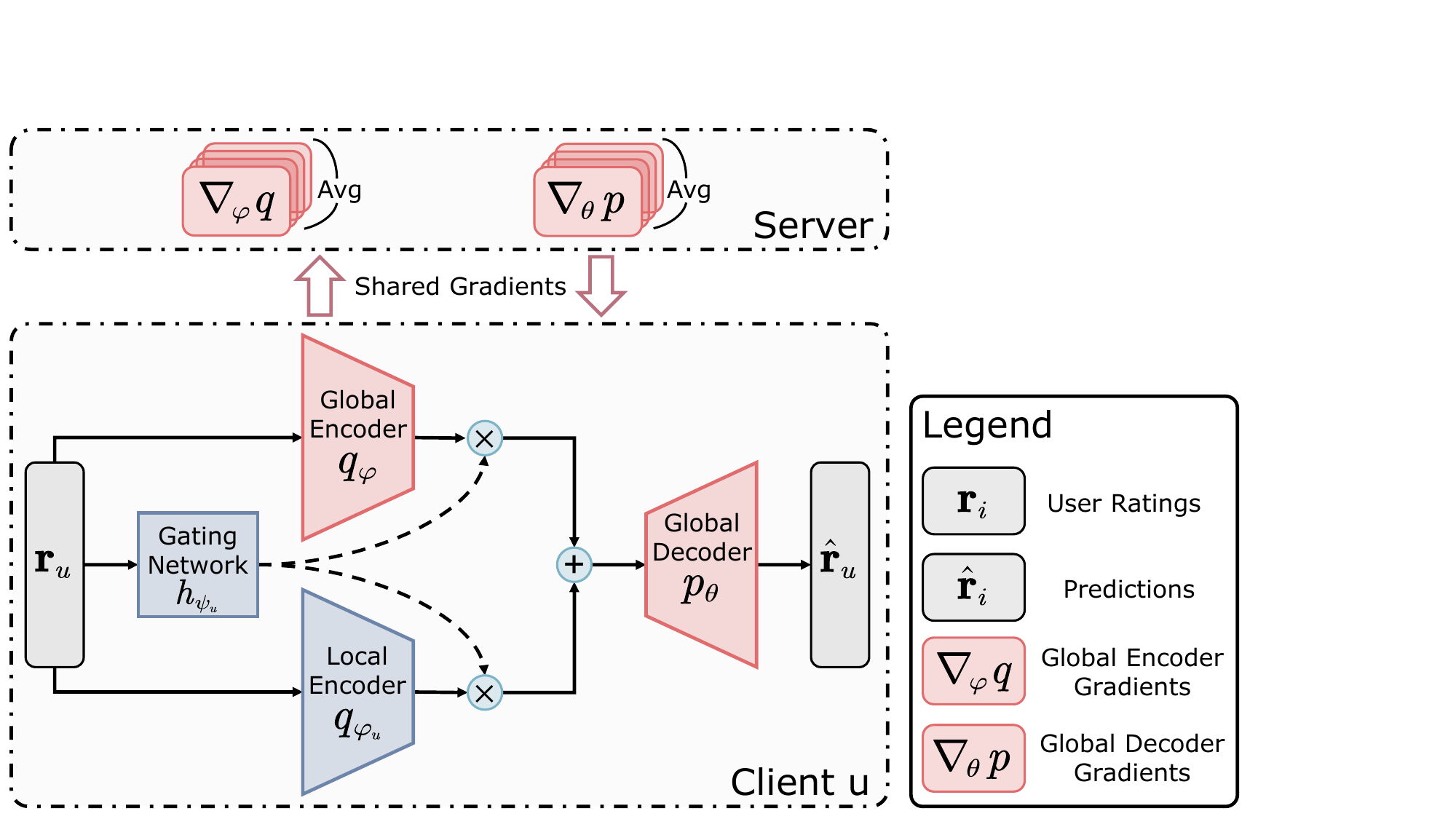}
    \caption{ The framework of FedDAE. }
    \label{fig:framework}
\end{figure}

\section{Methodology}

\subsection{Framework}
Fig. \ref{alg:algorithm} shows the overall framework of FedDAE. 
For each client \( u \), FedDAE inputs the interaction vector \(\mathbf{r}_u\) into the global encoder \(q_{\varphi}\), the local encoder \(q_{\varphi_u}\), and the gating network \(h_{\psi_u}\), separately. 
By multiplying the outputs of the dual encoders by the weights generated by \(h_{\psi_u}\) based on the user's interaction data and then summing them, FedDAE achieves adaptive additive personalization. 
The resulting output is then passed to the global decoder \(p_{\theta}\) to reconstruct the user's interaction data \(\hat{\mathbf{r}}_u\) \((u = 1,2,\dots,n)\). 
During the training phase, the global encoder and global decoder are collaboratively trained across clients, while the local encoder and gating network are trained locally.

\subsection{Dual Encoders}
%
We propose a dual-encoder mechanism to separately preserve shared knowledge across clients and client-specific personalized knowledge. 
Specifically, the encoder \( q_{\mathbf{\Phi}_u} \) in Eq. \ref{eq:combine_posterior} is implemented through a dual-encoder structure, consisting of a global encoder \( q_{\varphi} \) and a local encoder \( q_{\varphi_u} \) for each user \(u\). 
The parameters \(\varphi\) and \(\varphi_u\) capture the globally shared representation of item features and the personalized representation specific to user \(u\), respectively. 
This enables FedDAE to achieve personalized item feature representations while maintaining shared information.

\subsection{Gating Network}

To effectively combine the global and local representations, we use a gating network $h_{\psi_u}(\mathbf{r}_u) = [\omega_{u1}, \omega_{u2}]\in \mathbb{R}^2$, parameterized by $\psi_u$, which dynamically assigns weights $\omega_{u1}$ and $\omega_{u2}$ to the outputs of $g_{\varphi}$ and $g_{\varphi_u}$ based on the data $\mathbf{r}_u$. 

\begin{lemma}[Additivity of Gaussian distributions]
    \label{lemma:weight_sum_of_gaussian}
    Given two independent Gaussian random variables \(X\) and \(Y\), distributed as \(N(\mu_1, \sigma_1^2)\) and \(N(\mu_2, \sigma_2^2)\), respectively. 
    Then \(Z = w_1 X + w_2 Y\) follows a new Gaussian distribution \(Z \sim N(w_1 \mu_1 + w_2 \mu_2, w_1^2 \sigma_1^2 + w_2^2 \sigma_2^2)\).
\end{lemma}

Considering Lemma \ref{lemma:weight_sum_of_gaussian}, since the latent variable $\mathbf{z}_u$ is drawn from $q_{\mathbf{\Phi}_u} (\mathbf{z}_{u}| \mathbf{r}_{u})$ according to Eq. \eqref{eq:combine_posterior}, the combined \(\mu_{\mathbf{\Phi}_u}(\mathbf{r}_{u})\) and \(\sigma_{\mathbf{\Phi}_u}^{2}(\mathbf{r}_{u})\) are then defined as follows:
\begin{equation}
    \begin{split}
        \mu_{\mathbf{\Phi}_u}(\mathbf{r}_{u}) = \omega_{u1} \cdot \mu_{\varphi}(\mathbf{r}_{u}) + \omega_{u2} \cdot \mu_{\varphi_u}(\mathbf{r}_{u}), \\
        \sigma_{\mathbf{\Phi}_u}^{2}(\mathbf{r}_{u}) = \omega_{u1}^2 \cdot \sigma^2_{\varphi}(\mathbf{r}_{u}) + \omega_{u2}^2 \cdot \sigma^2_{\varphi_u}(\mathbf{r}_{u}),
    \end{split}
\end{equation}
where \(\mathbf{\Phi}_u = \{\varphi, \varphi_u, \psi_u\}\).
Through the gating network \(h_{\psi_u}\), FedDAE can adaptively adjust the balance between shared and personalized information based on each client's interaction data, thereby achieving personalized FedCF.

\subsection{Reconstruction and Prediction}
We define the VAE loss for the FedCF problem as follows. 
By combining Eq. \eqref{eq:condition_likelihood} and Eq. \eqref{eq:combine_posterior} to form a VAE, the evidence lower bound (ELBO) on the \(u\)-th client is given by:
\begin{equation}
    \resizebox{.9\linewidth}{!}{$
    \begin{split}
        & \mathcal{L}_{\beta}(\mathbf{r}_u; \mathbf{\Phi}_u, \theta) = \\
        & \mathbb{E}_{q_{\mathbf{\Phi}_u} (\mathbf{z}_{u}| \mathbf{r}_{u})}[\log p_{\mathbf{\theta}} (\mathbf{r}_{u} | \mathbf{z}_{u})]
        - \beta \cdot \operatorname{KL}(q_{\mathbf{\Phi}_u}(\mathbf{z}_{u}| \mathbf{r}_{u}) \| p_{\theta}(\mathbf{z}_u)),
    \end{split}
    $}
    \label{eq:obj_u}
\end{equation}
where \(\operatorname{KL}(\cdot)\) is the Kullback-Leibler divergence, and \(\beta \in [0, 1]\) is a hyperparameter controlling the strength of regularization, following the \(\beta\)-VAE \cite{higgins2017beta} .
The first term in Eq. \eqref{eq:obj_u} is the reconstruction loss, which equals to \(-\mathcal{L}_{recon}(\hat{\mathbf{r}}_u, \mathbf{r}_u)\), while the second term is for prior matching. 
By jointly maximizing Eq. (8) of all users, we obtain the reconstructed interactions \(\hat{\mathbf{r}}_u\) for each user \( u \) \((u=1,2,\dots,n)\).
However, relying solely on sharing  \(\theta\) is insufficient for effective information sharing across clients, which may result in suboptimal recommendation performance in a federated setting. 
To address this issue, we propose a novel personalized FedCF method called FedDAE.

After sampling \(\mathbf{z}_{u}\) from \(q_{\mathbf{\Phi}_u}(\mathbf{z}_{u}|\mathbf{r}_{u})\), it is decoded using the decoder \(p_{\theta}\) defined in Eq. \eqref{eq:condition_likelihood} to obtain the reconstructed interactions \(\hat{\mathbf{r}}_u\).
Combining Eq. \eqref{eq:obj_u}, the objective of FedDAE is to maximize the ELBO to approximateas \(\log p(\mathbf{r}_u)\) across all clients (\(u=1,2,\dots, n\)), which is defined as follows:
\begin{equation}
    \max_{\varphi, \{\varphi_u\}, \{\psi_u\}, \theta} 
    \sum_{u=1}^n \alpha_u
    \mathcal{L}_{\beta}(\mathbf{r}_u; \varphi, \varphi_u, \psi_u, \theta).
    \label{eq:obj}
\end{equation}
Here, \(\alpha_u\) is the weight of the loss for the \(u\)-th client, used to balance the contribution of this client, satisfying \(\sum_{u=1}^{n}\alpha_u=1\).
Once the objective function in Eq. \eqref{eq:obj} is optimized, we obtain the final prediction \(\hat{\mathbf{r}}_u = \mu_{\mathbf{\Phi}_u}(\mathbf{r}_{u})\), which will be used to personalize recommendations for the \( u \)-th user, suggesting items they might be interested in.

\subsection{Algorithm}
To optimize the objective function outlined in Eq. \eqref{eq:obj}, we utilize an alternative optimization algorithm for training the FedDAE method.
The overall workflow of this algorithm is summarized in several steps, as demonstrated in the Alg. \ref{alg:algorithm}.

The process begin by initializing \(\varphi\) and \(\theta\) on the server, and all \(\varphi_u\) and \(\psi_u\) on their respective clients. 
For each communication round \(t\), the server randomly selects a subset of clients to participate in the training, denoted as \(S_t\). 
It is important to note that, as discussed in \cite{chai2020secure,li2024federated}, for privacy protection, no client should participate in training for two consecutive rounds. 
According to the line 4 in the \(GlobalProcedure\), the latent variable \(\mathbf{z}_u\) is sampled from the normal Gaussian distribution using the reparameterization trick.
After receiving \(\varphi\) and \(\theta\), the selected clients call the \(ClientUpdate\) function to update their local models with the learning rate \(\eta\).

In the \(ClientUpdate\) function, The local model updates are performed over \(E\) epochs (line 13).
Once the update is completed, the accumulated gradients \(\nabla_{\varphi}^{(u)}\) and \(\nabla^{(u)}_{\theta}\), and the predicted scores \(\hat{\mathbf{r}}_u\) are uploaded to the server for global aggregation.
The server then apply the updated \(\varphi\) and \(\theta\) to the next training round. After the training phase is completed, FedDAE use the predicted scores \(\hat{\mathbf{R}}\) as the recommendation guide.
On each client, FedDAE constructs a global encoder to model the global representation of item features and a local encoder to capture the personalized representation of features influenced by user preferences. 
The outputs of these encoders are weighted and combined through a gating network, achieving adaptive additive personalization in recommendations. 
Therefore, despite variations in users' interaction data due to their preferences, FedDAE ensures that data within a user adheres to the i.i.d. assumption. 
Additionally, the user-related local encoder is stored only locally on the client, protecting user privacy.

\begin{algorithm}[!htbp]
    \begin{minipage}{1\linewidth}
    \caption{FedDAE}
    \label{alg:algorithm}
    
    \hspace*{0.05in}{\bf Input}: \(\mathbf{R}\), \(\beta\), \(k\), \(\eta\), \(T\), \(E\) \\
    \hspace*{0.05in}{\bf Initialize}: \(\varphi\), \(\{\varphi_u\}\), \(\{\psi_u\}\), \(\theta\) \\
    \hspace*{0.05in}{\bf GlobalProcedure}:
    \begin{algorithmic}[1]

        \FOR{\(t=1,2,\dots,T\)}
            \STATE \(S_t \leftarrow\) randomly select \(n_s\) from \(n\) clients
            \FORALL{client index \(u \in S_t\)}
                \STATE Sample \(\epsilon \sim \mathcal{N}(0, \mathbf{I}_k)\) and compute \(\mathbf{z}_u\) via reparameterization trick;
                \STATE \(\hat{\mathbf{r}}_u\), \(\nabla_{\varphi}^{(u)}\), \(\nabla^{(u)}_{\theta} \leftarrow \text{ClientUpdate}(\varphi, \theta)\);
            \ENDFOR
            
            \STATE \(\varphi = \varphi - \frac{\eta}{n_s} \sum_{u \in S_t} \nabla_{\varphi}^{(u)}\); 
            \STATE \(\theta = \theta - \frac{\eta}{n_s} \sum_{u \in S_t} \nabla^{(u)}_{\theta}\); 

        \ENDFOR

        \STATE \bf return: \(\hat{\mathbf{R}} = [\hat{r}_1, \hat{r}_2, \dots, \hat{r}_n]^T\)
    \end{algorithmic}

    \hspace*{0.05in}{\bf ClientUpdate}:
    \begin{algorithmic}[1]
        \STATE \(\nabla_{\varphi} = 0\), \(\nabla_{\theta}=0\);
        \FOR{\(e=1,2,\dots,E\)} 
            \STATE Compute gradients \(\nabla_{\varphi} \mathcal{L}_{\beta}\), \(\nabla_{\varphi_u} \mathcal{L}_{\beta}\), \(\nabla_{\psi_u} \mathcal{L}_{\beta}\), \(\nabla_{\theta} \mathcal{L}_{\beta}\) according to Eq. \eqref{eq:obj};
            \STATE Use gradient descent to update \(\varphi\), \(\varphi_u\), \(\psi_u\), \(\theta\) with \(\eta\);
            \STATE \(\nabla_{\varphi} = \nabla_{\varphi} + \nabla_{\varphi} \mathcal{L}_{\beta}\);
            \STATE \(\nabla_{\theta} = \nabla_{\theta} + \nabla_{\theta} \mathcal{L}_{\beta}\);
            \STATE Compute \(\hat{\mathbf{r}}_u\);
        \ENDFOR

        \STATE \bf return: \(\hat{\mathbf{r}}_u\), \(\nabla_{\varphi}\), \(\nabla_{\theta}\)
    \end{algorithmic}
    \end{minipage}
\end{algorithm}

\section{Discussions}
\subsection{Matrix Factorization}
In recommendation systems, MF-based CF methods typically decompose the interaction data matrix into an item embedding that preserves shared knowledge and a user embedding that retains personalized knowledge. 
In our proposed method, we model the data distribution of the interaction matrix and use the weights of the dual encoders to capture both the globally shared knowledge across clients and the personalized knowledge for each user. 
This allows our method, FedDAE, to retain more complex information, rather than being limited to the form of user-item vectors.

\subsection{Complexity Analysis}
Given $n$ users and $m$ items, the input dimension of the encoder is $m$. Assuming each encoder on the client side consists of $L$ fully connected layers with a hidden layer dimension of $d >> k$, the time complexity of the encoder is $O(Lmd)$, and the space complexity is $O(L(m + 1)d)$. 
Similarly, assuming the decoder has $L'$ fully connected layers with the same hidden layer dimension $d$, the time and space complexities of the decoder are $O(L'md)$ and $O(L'(m + 1)d)$ respectively. 
The reparameterization involves generating the mean and variance of the latent subspace $\mathbf{z}_u$ and sampling using a standard normal distribution, with a time complexity of $O(k)$. 
Additionally, the gating network on each client requires $O(2m)$ time and space complexity.
Thus, the overall time complexity is $O(n(2L + L')md)$. 
Due to each client in FedDAE having an independent gating network, an independent encoder, a globally shared encoder, and a globally shared decoder, the overall space complexity is $O(((n+1)L + L')(m + 1)d)$.

\subsection{Privacy-preservation Enhancement}
The proposed framework adopts a decentralized architecture from the FL scheme, which significantly reduces the risk of privacy leaks by maintaining data locality, which generally performs well in a trusted environment. 
Similar to most FedCF methods \cite{chai2020secure,zhang2023dual,li2024federated}, our approach only transmits the gradients of global model parameters, \(\nabla_{\varphi}^{(u)} \mathcal{L}_{\beta}\) and \(\nabla^{(u)}_{\theta} p\), without sharing any raw data with third parties.
Additionally, if in an untrusted environment, the proposed method can be easily combined with other 
advanced privacy-preserving techniques, such as differential privacy \cite{abadi2016deep} or randomly cropping gradients,
to further strengthen user privacy guarantees.
In our experiment, we perform an ablation study to evaluate the effectiveness of privacy protection.

\section{Experiments}

\begin{table}[!htbp]
\centering
    \resizebox{.8\linewidth}{!}{
    \begin{tabular}{lcccc}
        \hline
        Datasets    & \#Ratings   & \#Users  & \#Items  & Sparsity \\ \hline
        ML-100k     & 100,000     & 943      & 1,682    & 93.70\%  \\
        ML-1M       & 1,000,209   & 6,040    & 3,706    & 95.53\%  \\
        Video       & 23,181      & 1,372    & 7,957    & 99.79\%  \\
        QB-article  & 266,356     & 24,516   & 7,455    & 99.81\%  \\ \hline
    \end{tabular}
    }
    \caption{
        The statistical information of the used datasets.
    }
    \label{table:datasets}
\end{table}

\subsection{Datasets}
We perform an extensive experimental analysis to assess the performance of FedDAE on four widely utilized datasets:
MovieLens-100K (ML-100K), MovieLens-1M (ML-1M), Amazon-Instant-Video (Video), and QB-article. 
Table \ref{table:datasets} provides the statistical details of the datasets. 
Each dataset exhibits a high degree of sparsity, with the percentage of observed ratings compared to the total possible ratings (\#Users $\times$ \#Items) exceeding 90\%.
The first three datasets contain explicit ratings ranging from 1 to 5. 
Since FedDAE focuses on generating recommendation predictions for data with implicit feedback, any rating above 0 in these datasets is considered a positive interaction by the user and is assigned 1. 
The QB-article dataset is an implicit feedback dataset that records user click behavior. In each dataset, we only include users who rated at least 10 items.
All datasets used are publicly available, and details are provided in the appendix.


\subsection{Baselines}
The efficacy of FedDAE is comparatively evaluated against several cutting-edge approaches in both centralized and federated environments for validation:

\noindent\textbf{Mult-VAE} \cite{liang2018variational}: A variational autoencoder model that uses a multinomial distribution as the likelihood function to generate latent representations of users, suitable for collaborative filtering tasks with implicit feedback data in recommendation systems.

\noindent\textbf{RecVAE} \cite{shenbin2020recvae}: significantly improves the performance of Mult-VAE by introducing a composite prior distribution and an alternating training method,

\noindent\textbf{LightGCN} \cite{he2020lightgcn}: enhances recommendation performance to a new level while retaining the advantages of GCN models by simplifying the design of graph convolutional networks (GCNs) and removing nonlinear transformations and weight matrices, 

\noindent\textbf{FedVAE} \cite{polato2021federated}: extends Mult-VAE to the FL framework, achieving collaborative filtering by only transmitting model gradients between clients and the server without exposing the raw user data, ensuring user privacy and security.

\noindent\textbf{PFedRec} \cite{zhang2023dual}: achieves the personalization of user information and item features by introducing a dual personalization in the FL framework, enabling the recommendation system to adapt to the needs of different users.

\noindent\textbf{FedRAP} \cite{li2024federated}: enhances the performance of recommendation systems by applying dual personalization of user and item information in federated learning while retaining the shared parts of item information, particularly excelling in handling heterogeneous data.

The implementations of these baselines are publicly available in their respective papers.
Additionally, we implement a central version of FedDAE, named \textbf{CentDAE}, to explore the performance of a VAE with dual encoders.

\begin{table*}[!htbp]
\centering
\resizebox{1\textwidth}{!}{
\begin{tabular}{clcccccccc}
\hline
\multirow{2}{*}{}          & \multirow{2}{*}{Method} & \multicolumn{2}{c}{ML-100K}                                 & \multicolumn{2}{c}{ML-1M}                                   & \multicolumn{2}{c}{Video}                                      & \multicolumn{2}{c}{QB-Article}                              \\
                           &                         & HR@20                        & NDCG@20                      & HR@20                        & NDCG@20                      & HR@20                           & NDCG@20                      & HR@20                        & NDCG@20                      \\ \hline
\multirow{4}{*}{Central}   & MultiVAE                & 0.1093 $\pm$ 0.0143          & 0.0436 $\pm$ 0.0065          & 0.0682 $\pm$ 0.0016          & 0.0276 $\pm$ 0.0005          & 0.0689    $\pm$ 0.0088          & 0.0245 $\pm$ 0.0028          & 0.0129 $\pm$ 0.0025          & 0.0061 $\pm$ 0.0008          \\
                           & RecVAE                  & 0.1526 $\pm$ 0.0046          & 0.0590 $\pm$ 0.0017          & 0.1067 $\pm$ 0.0029          & 0.0409 $\pm$ 0.0010          & 0.0693  $\pm$   0.0041          & 0.0257 $\pm$ 0.0028          & 0.0195 $\pm$ 0.0075          & 0.0064 $\pm$ 0.0020          \\
                           & LightGCN                & 0.1761 $\pm$ 0.0166          & 0.0953 $\pm$ 0.0128          & 0.1207 $\pm$ 0.0015          & 0.0677 $\pm$ 0.0014          & 0.0761  $\pm$   0.0056          & 0.0323 $\pm$ 0.0013          & 0.0204 $\pm$ 0.0067          & 0.0070 $\pm$ 0.0039          \\
                           & CentDAE                 & 0.1119 $\pm$ 0.0024          & 0.0453 $\pm$ 0.0010          & 0.0702 $\pm$ 0.0008          & 0.0275 $\pm$ 0.0004          & 0.0702  $\pm$   0.0084          & 0.0254 $\pm$ 0.0034          & 0.0185 $\pm$ 0.0033          & 0.0064 $\pm$ 0.0007          \\ \hline
\multirow{4}{*}{Federated} & FedVAE                  & 0.1200 $\pm$ 0.0040          & 0.0501 $\pm$ 0.0032          & 0.0639 $\pm$ 0.0012          & 0.0250 $\pm$ 0.0006          & 0.0634    $\pm$ 0.0128          & 0.0227 $\pm$ 0.0036          & 0.0116 $\pm$ 0.0031          & 0.0045 $\pm$ 0.0013          \\
                           & PFedRec                 & 0.0541 $\pm$ 0.0054          & 0.0115 $\pm$ 0.0026          & 0.0600 $\pm$ 0.0002          & 0.0231 $\pm$ 0.0000          & 0.0138  $\pm$   0.0064          & 0.0026 $\pm$ 0.0012          & 0.0065 $\pm$ 0.0002          & 0.0033 $\pm$ 0.0000          \\
                           & FedRAP                  & 0.0626 $\pm$ 0.0032          & 0.0122 $\pm$ 0.0017          & 0.0625 $\pm$ 0.0015          & 0.0240 $\pm$ 0.0003          & 0.0142  $\pm$   0.0097          & 0.0028 $\pm$ 0.0019          & 0.0116 $\pm$ 0.0026          & 0.0035 $\pm$ 0.0010          \\
                           & FedDAE                  & \textbf{0.1232 $\pm$ 0.0023} & \textbf{0.0503 $\pm$ 0.0014} & \textbf{0.0650 $\pm$ 0.0039} & \textbf{0.0261 $\pm$ 0.0011} & \textbf{0.0654  $\pm$   0.0080} & \textbf{0.0230 $\pm$ 0.0029} & \textbf{0.0124 $\pm$ 0.0030} & \textbf{0.0052 $\pm$ 0.0014} \\ \hline
\end{tabular}
}
\caption{
       Experimental results on HR@20 and NDCG@20 shown in percentages on four real-world datasets. 
       \textbf{Central} and \textbf{Federated} represent centralized and federated methods, respectively. 
       The best results are highlighted in boldface.
}
\label{table:main_results}
\end{table*}

\subsection{Experimental Setting}

We refer to previous work \cite{he2017neural,zhang2023dual,li2024federated}, randomly selecting 4 negative samples for each positive sample in the training set, and using the leave-one-out strategy to validate the methods.
In this work, we set $h_{\psi_u}(\mathbf{r}_u) = \operatorname{softmax}(\mathbf{r}_u \cdot \psi_u)$ and thus $w_{u1}+w_{u2}=1$.
We perform hyper-parameter tuning for FedDAE and select the learning rate $\eta$ from $\{10^i | i = -8, ..., -1\}$.
The Adam optimizer \cite{kingma2014adam} is applied to update FedDAE's parameters.
Referring to previous work \cite{liang2018variational,polato2021federated}, we gradually anneal $\beta=1$.
To ensure fairness, we fix the latent embedding dimension to 256 and the training batch size to 2048 for all methods. The number of layers for methods with hidden layers is fixed at 3. 
We set $\alpha_u = 1/n$ for the average schemes of all userd federated methods.
Following Mult-VAE, FedDAE also employs a dropout \cite{srivastava2014dropout} layer before inputting $\mathbf{r}_u$ into the encoders to reduce the risk of over-fitting and uses the reparametrization trick \cite{kingma2013auto} to eliminate the stochasticity caused by sampling $\mathbf{z}_u$, which allows the model to be optimized using gradient descent.
In the main experiments, all clients participated in the training for FedVAE, PFedRec, FedRAP, and FedDAE, and no additional privacy protection measures are applied. In addition, the number of local epoch is set to 10 for these federated methods, and we do not use any pre-training strategies for any methods.

\subsection{Evaluation}
We evaluate the prediction performance in the experiments using two widely used metrics: \textit{Hit Rate} (HR@K) and \textit{Normalized Discounted Cumulative Gain} (NDCG@K). These criteria were formally defined in the work \cite{he2015trirank}. 
HR measures hit rate to show whether relevant items are recommended, while NDCG accounts for ranking quality, emphasizing user experience by ranking items of greater interest higher.
In this study, we set K = 20 and repeat all experiments five times to report the mean and standard deviations.

\begin{figure}[!htbp]
    \centering
    \includegraphics[width=.99\linewidth]{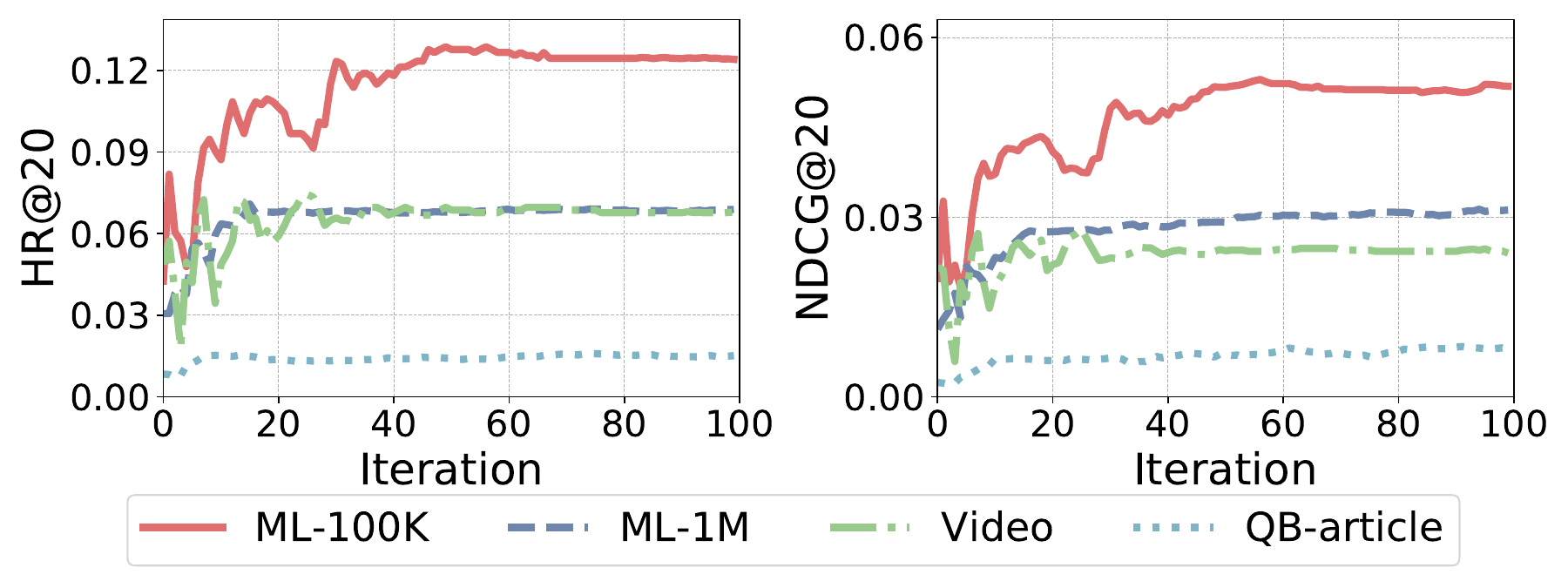}
    \caption{
        The visualization of FedDAE's convergence and efficiency across the four used datasets.
    }
    \label{fig:convergence}
\end{figure}

\subsection{Comparison Analysis}

Table \ref{table:main_results} presents the experimental results of all baseline methods and FedDAE on four widely used recommendation datasets, demonstrating that FedDAE outperforms all federated methods and achieves performance close to centralized methods. 
This is attributed to FedDAE's ability to adaptively and individually model each user's data distribution, retaining shared information about item features while better integrating user-specific item representations. This indicates that FedDAE has learned fine-grained personalized features, effectively leveraging user preferences.

Additionally, to study FedDAE's convergence, we evaluated its performance over 100 iterations on all datasets and visualized the HR@20 and NDCG@20 results, as shown in  Fig. \ref{fig:convergence}. Insights from Table \ref{table:main_results} and Fig. \ref{fig:convergence} reveal that the sparsity and scale of the datasets significantly affect FedDAE's performance. 
On datasets with lower sparsity and moderate scale, such as ML-100K and ML-1M, FedDAE shows rapid convergence and better results. 
In contrast, on datasets with higher sparsity, like Video and QB-article, the model's convergence speed and final performance are noticeably slower. This indicates that additional strategies may be needed to improve the model's learning ability when dealing with highly sparse datasets. 
Furthermore, while the Adam optimizer helps quickly find optimal model parameters and enhances recommendation performance, it also causes more noticeable fluctuations in FedDAE's convergence curves. These fluctuations might be due to the heterogeneity and noise in the datasets.
A theoretical convergence analysis of the convergence is provided in our appendix.

\subsection{Ablation Study}

To investigate the impact of each component on the model's performance, we propose a variant of the FedDAE method, named \textbf{FedDAE$_{w}$}.
It uses a fixed weight ww for each client to weigh the output of the global encoder. Since the weight of the local encoder is complementary to \(w\) (i.e., \(1 - w\)), the local encoder's weight is also fixed. 
For more ablation studies, please refer to our appendix.


\begin{table}[!htbp]
\centering
\resizebox{1\linewidth}{!}{
\begin{tabular}{ccccc}
\hline
        & FedDAE & FedDAE\(_{w=0.25}\)            & FedDAE\(_{w=0.5}\)            & FedDAE\(_{w=0.75}\)             \\ \hline
HR@20   & 0.1287 & 0.1213 (5.75\%\(\downarrow\))  & 0.1245 (3.26\%\(\downarrow\)) & 0.1191 (7.46\%\(\downarrow\))   \\
NDCG@20 & 0.0524 & 0.0506 (3.44\%\(\downarrow\))  & 0.0513 (2.10\%\(\downarrow\)) & 0.0469 (10.50\%\(\downarrow\))  \\ \hline
\end{tabular}
}
\caption{
    Performance Comparison of FedDAE and its fixed weight variants FedDAE\(_w\) (\(w=0.25, 0.5, 0.75\)) on the ML-100K dataset.
    The values in parentheses indicate the percentage decrease compared to the FedDAE method.
}
\label{table:global_local}
\end{table}

\paragraph{Abalation Study on Adaptive Weights.}
To explore the impact of the gating network output weight on each client in FedDAE, we compared the performance of FedDAE$_w$ with different fixed weights ($w = 0.25, 0.5, 0.75$) and FedDAE on the ML-100K dataset. Table 3 shows that when the weight $w = 0.5$, FedDAE$_w$ performs most similarly to FedDAE. However, as the weight increases to 0.75, performance significantly decreases, indicating that personalized information plays a crucial role in FedDAE. The lack of personalized information results in a greater decline in recommendation performance compared to the lack of shared information. Overall, this ablation study demonstrates the effectiveness of the gating network in FedDAE.

\begin{figure}[!t]
    \centering
    \includegraphics[width=1\linewidth]{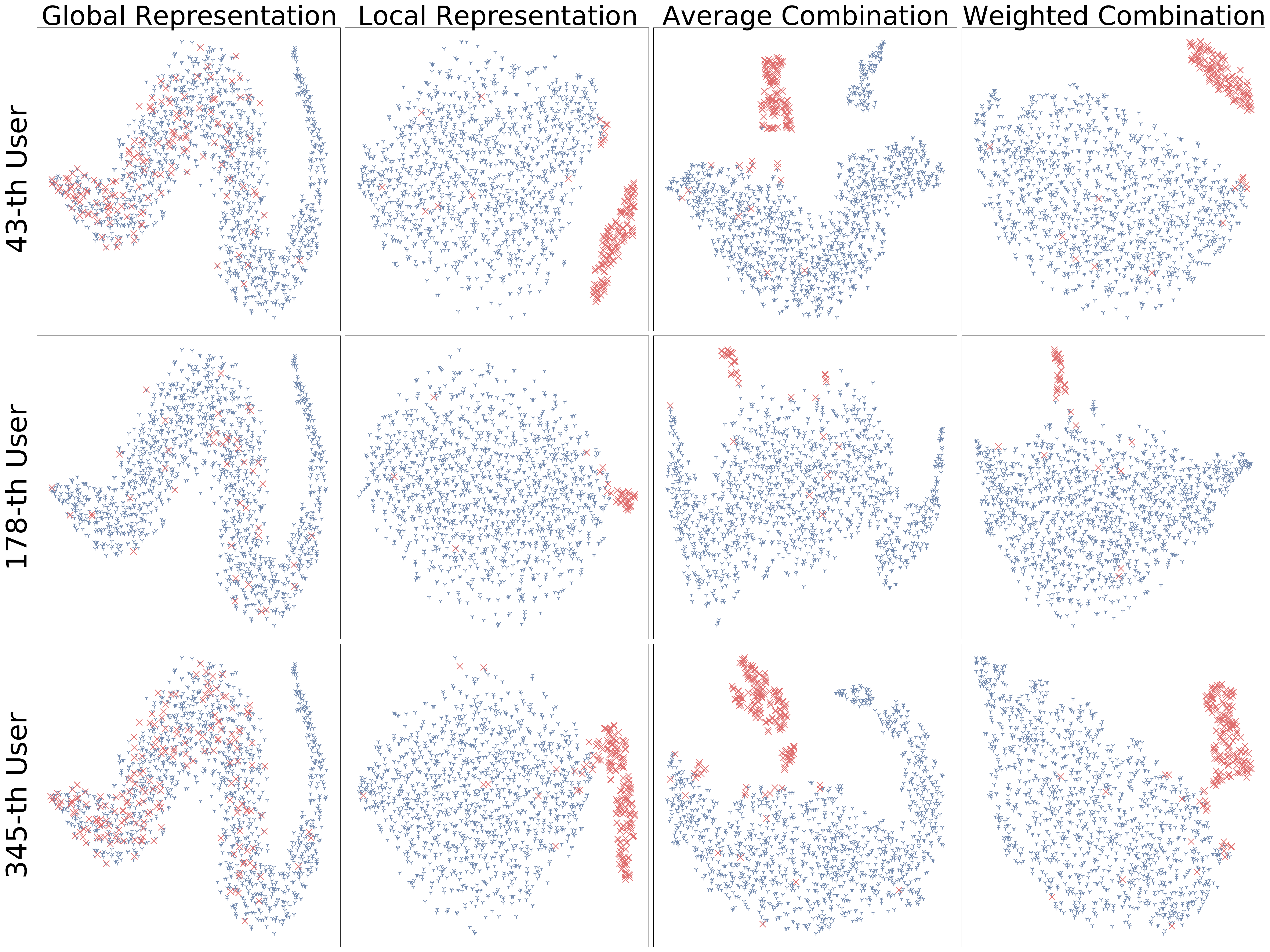}
    \caption{
        The t-SNE visualization of item features learned by FedDAE on the ML-100K dataset illustrates the representations across different users.
        In the visualization, red indicates items that users have interacted with, while blue indicates items they have not interacted with. 
        The global representation is generated by FedDAE's global encoder, and the local representation is produced by the user-specific encoder.
        \textbf{Average Combination} refers to the simple average of the global and local representations, while \textbf{Weighted Combination} reflects the weighted combination based on the gating network outputs tailored to the client data. 
        FedDAE's adaptive personalization enhances its ability to distinguish between items users have interacted with and those they haven't, leading to improved recommendation performance.
    }
    \label{fig:item_representations}
\end{figure}

\paragraph{Ablation Study on Item Representation.}
To visually demonstrate the differentiated capability of FedDAE encoders in modeling item features, we selected three local encoders $\{q_{\varphi_u}\}_{u=43, 178, 345}$ from different users and the global encoder $q_{\varphi}$, all learned from the ML-100K dataset. 
The weights of all layers of each encoder were multiplied and then visualized. 
Since this study primarily focuses on implicit feedback recommendation, where each item is either interacted with by the user or not, We employe t-SNE \cite{van2008visualizing} to map these item embeddings into a two-dimensional space, with the normalized results shown in Fig. \ref{fig:item_representations}. In these visualizations, blue represents items not interacted with by the user, while red represents items that have been interacted with.
From the first column of Fig. \ref{fig:item_representations}, we can see that in the global encoder’s modeled item representation, interacted and non-interacted items are mixed together, indicating that $q_{\varphi}$ only models the shared features of items. 
The second column illustrates that each user's local encoder can divide the item features into two distinct clusters, proving that the local encoder $q_{\varphi_u}$ can learn user-specific personalized features.
The third column shows the average combination of global and local representations, but its boundaries are not as distinct compared to the weighted combination of FedDAE shown in the fourth column. 
The weighted combination, based on the output of the gating network $h_{\psi}$ from user interaction data $\mathbf{r}_u$, more clearly differentiates the two clusters. 
This demonstrates that FedDAE’s adaptive additive personalization can more effectively represent item features, supporting its superior performance in personalized recommendation.

\begin{table}[!htbp]
\centering
\resizebox{1\linewidth}{!}{
\begin{tabular}{clcccc}
\hline
$\kappa^2$         & Metric  & FedDAE & FedDAE$_{w=0.25}$           & FedDAE$_{w=0.5}$            & FedDAE$_{w=0.75}$           \\ \hline
\multirow{2}{*}{0.2} & HR@20   & 0.1255 & 0.1213 (3.35\%$\downarrow$) & 0.1245 (0.80\%$\downarrow$) & 0.1149 (8.45\%$\downarrow$) \\
                     & NDCG@20 & 0.0505 & 0.0483 (4.36\%$\downarrow$) & 0.0486 (3.76\%$\downarrow$) & 0.0476 (5.74\%$\downarrow$) \\
\multirow{2}{*}{0.5} & HR@20   & 0.1245 & 0.1181 (5.14\%$\downarrow$) & 0.1213 (2.57\%$\downarrow$) & 0.1170 (6.02\%$\downarrow$) \\
                     & NDCG@20 & 0.0494 & 0.0479 (3.04\%$\downarrow$) & 0.0481 (2.63\%$\downarrow$) & 0.0470 (4.86\%$\downarrow$) \\
\multirow{2}{*}{0.8} & HR@20   & 0.1191 & 0.1170 (1.76\%$\downarrow$) & 0.1181 (0.84\%$\downarrow$) & 0.1160 (2.60\%$\downarrow$) \\
                     & NDCG@20 & 0.0489 & 0.0473 (3.27\%$\downarrow$) & 0.0486 (0.61\%$\downarrow$) & 0.0464 (5.11\%$\downarrow$) \\
\multirow{2}{*}{1}   & HR@20   & 0.1191 & 0.1085 (8.90\%$\downarrow$) & 0.1096 (7.98\%$\downarrow$) & 0.1085 (8.90\%$\downarrow$) \\
                     & NDCG@20 & 0.0479 & 0.0453 (5.43\%$\downarrow$) & 0.0466 (2.71\%$\downarrow$) & 0.0450 (6.05\%$\downarrow$) \\ \hline
\end{tabular}
}
\caption{
    Performance comparison of FedDAE and FedDAE\(_w\) variants under Different Noise Variance Levels on the ML-100K Dataset, showing the performance metrics HR@20 and NDCG@20 for FedDAE and its variants with fixed weights (\(w=0.25, 0.5, 0.75\)) under different noise variance levels (0.2, 0.5, 0.8, 1). The values in parentheses indicate the percentage decrease compared to FedDAE. 
}
\label{table:noisy}
\end{table}

\paragraph{Abalation Study on Privacy Protection.}

To evaluate the impact of additional privacy protection measures on the recommendation performance of FedDAE and FedDAE\(_w\), we introduce noise into the gradients uploaded by users and test these methods on the ML-100K dataset across various noise levels (\(\kappa^2=0, 0.2, 0.5, 0.8, 1\)). 
Table \ref{table:noisy} shows that while performance declines as noise increases, FedDAE consistently outperforms other variants at all noise levels, likely due to its adaptive balancing of global and local encoder outputs. 
FedDAE\(_{w=0.5}\) ranks second, demonstrating strong robustness across different noise levels. 
In contrast, FedDAE\(_{w=0.25}\) performs worse than both FedDAE and FedDAE\(_{w=0.5}\) but better than FedDAE\(_{w=0.75}\). As the weight parameter \(w\) increases, FedDAE becomes more reliant on the global encoder, which may explain why FedDAE\(_{w=0.75}\) performs the worst under all noise conditions. 
These results highlight the importance of balancing the weights of the dual encoders, especially when implementing privacy protections.

\section{Conclusion}

This paper first revisits FedCF from the perspective of VAEs and proposes a novel personalized FedCF method called FedDAE. 
FedDAE constructs a VAE model with dual encoders and a global decoder for each client, capturing user-specific representations of item features while preserving globally shared information. 
The outputs of the two encoders are dynamically weighted through a gating network based on user interaction data, achieving adaptive additive personalization. 
Experimental results on four widely used recommendation datasets demonstrate that FedDAE outperforms existing FedCF methods and various ablation baselines, showcasing its ability to provide efficient personalized recommendations while protecting user privacy. 
Additionally, our research reveals that the sparsity and scale of the datasets significantly impact the performance of FedDAE, suggesting that highly sparse datasets require additional strategies to enhance the model's learning capability.


\bibliography{aaai25}

\section{Reproducibility Checklist}

\noindent Unless specified otherwise, please answer “yes” to each question if the relevant information is described either in the paper itself or in a technical appendix with an explicit reference from the main paper. If you wish to explain an answer further, please do so in a section titled “Reproducibility Checklist” at the end of the technical appendix.

\noindent This paper:
\begin{itemize}
    \item Includes a conceptual outline and/or pseudocode description of AI methods introduced (\underline{\textbf{yes}})
    \item Clearly delineates statements that are opinions, hypothesis, and speculation from objective facts and results (\underline{\textbf{yes}})
    \item Provides well marked pedagogical references for less-familiar readers to gain background necessary to replicate the paper (\underline{\textbf{yes}})
\end{itemize}

\noindent Does this paper make theoretical contributions? (\underline{\textbf{yes}})

\noindent If yes, please complete the list below.
\begin{itemize}
    \item All assumptions and restrictions are stated clearly and formally. (\underline{\textbf{yes}})
    \item All novel claims are stated formally (e.g., in theorem statements). (\underline{\textbf{yes}})
    \item Proofs of all novel claims are included. (\underline{\textbf{yes}})
    \item Proof sketches or intuitions are given for complex and/or novel results. (\underline{\textbf{yes}})
    \item Appropriate citations to theoretical tools used are given. (\underline{\textbf{yes}})
    \item All theoretical claims are demonstrated empirically to hold. (\underline{\textbf{yes}})
    \item All experimental code used to eliminate or disprove claims is included. (\underline{\textbf{yes}})
\end{itemize}

\noindent Does this paper rely on one or more datasets? (\underline{\textbf{yes}})

\noindent If yes, please complete the list below.
\begin{itemize}
    \item A motivation is given for why the experiments are conducted on the selected datasets (\underline{\textbf{yes}})
    \item All novel datasets introduced in this paper are included in a data appendix. (\underline{\textbf{NA}})
    \item All novel datasets introduced in this paper will be made publicly available upon publication of the paper with a license that allows free usage for research purposes. (\underline{\textbf{NA}})
    \item All datasets drawn from the existing literature (potentially including authors’ own previously published work) are accompanied by appropriate citations. (\underline{\textbf{yes}})
    \item All datasets drawn from the existing literature (potentially including authors’ own previously published work) are publicly available. (\underline{\textbf{yes}})
    \item All datasets that are not publicly available are described in detail, with explanation why publicly available alternatives are not scientifically satisfying. (\underline{\textbf{NA}})
\end{itemize}

\noindent Does this paper include computational experiments? (\underline{\textbf{yes}})

\noindent If yes, please complete the list below.
\begin{itemize}
    \item Any code required for pre-processing data is included in the appendix. (\underline{\textbf{yes}}).
    \item All source code required for conducting and analyzing the experiments is included in a code appendix. (\underline{\textbf{yes}})
    \item All source code required for conducting and analyzing the experiments will be made publicly available upon publication of the paper with a license that allows free usage for research purposes. (\underline{\textbf{yes}})
    \item All source code implementing new methods have comments detailing the implementation, with references to the paper where each step comes from (\underline{\textbf{partial}})
    \item If an algorithm depends on randomness, then the method used for setting seeds is described in a way sufficient to allow replication of results. (\underline{\textbf{yes}})
    \item This paper specifies the computing infrastructure used for running experiments (hardware and software), including GPU/CPU models; amount of memory; operating system; names and versions of relevant software libraries and frameworks. (\underline{\textbf{yes}})
    \item This paper formally describes evaluation metrics used and explains the motivation for choosing these metrics. (\underline{\textbf{yes}})
    \item This paper states the number of algorithm runs used to compute each reported result. (\underline{\textbf{yes}})
    \item Analysis of experiments goes beyond single-dimensional summaries of performance (e.g., average; median) to include measures of variation, confidence, or other distributional information. (\underline{\textbf{yes}})
    \item The significance of any improvement or decrease in performance is judged using appropriate statistical tests (e.g., Wilcoxon signed-rank). (yes/partial/no)
    \item This paper lists all final (hyper-)parameters used for each model/algorithm in the paper’s experiments. (\underline{\textbf{no}})
    \item This paper states the number and range of values tried per (hyper-) parameter during development of the paper, along with the criterion used for selecting the final parameter setting. (\underline{\textbf{yes}})
\end{itemize}

\end{document}


\maketitle

\renewcommand{\theequation}{A\arabic{equation}}
\renewcommand{\thefigure}{A\arabic{figure}}
\renewcommand{\thetable}{A\arabic{table}}

\newtheorem{assumption}{Assumption}
\newtheorem{theorem}{Theorem}
\newtheorem{discus}{Discussion}
\newtheorem{lemma}{Lemma}
\newtheorem{proof}{Proof}

\section{Appendix}

\subsection{Method Discussion}

\begin{figure}[!htbp]
    \centering
    \includegraphics[width=0.95\linewidth]{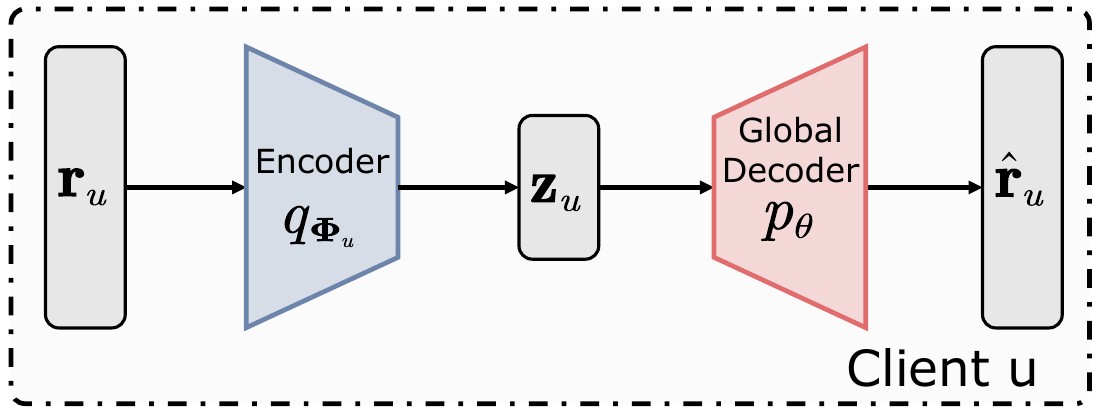}
    \caption{The abstract architecture of FedDAE on the \( u \)-th client. When the encoding part is abstracted into a single encoder, the model structure of FedDAE on the \( u \)-th client is similar to that of Mult-VAE.}
    \label{fig:ab_frame}
\end{figure}

In the Methods section, we introduced that FedDAE constructs a VAE with dual encoders on each client and combines the outputs of these encoders weighted according to the user's data on the client. Therefore, as shown in Fig. \ref{fig:ab_frame}, let us first define the encoding part of FedDAE as an abstract encoder \( q \), driven by three variables: \(\varphi\), \(\varphi_u\), and \(\psi\), collectively represented as \(\mathbf{\Phi}_u\).
Therefore, we have:
\begin{equation}
    \mathbf{z}_u \sim q_{\mathbf{\Phi}_u} (\mathbf{z}_u | \mathbf{r}_u), 
\end{equation}
which provides the theoretical basis for the construction of Eq. (5).

Now, revisiting Figure 2, we see that the latent variable \(\mathbf{z}_u\) on each client is determined by the outputs of both the global encoder \(q_{\varphi}\) and the local encoder \(q_{\varphi_u}\). Let \(\mathbf{z}^{(u)}_g \sim q_{\varphi}(\mathbf{z}^{(u)}_g|\mathbf{r}_u)\) represent the output of the global encoder \(q_{\varphi}\) and \(\mathbf{z}^{(u)}_l \sim q_{\varphi_u}(\mathbf{z}^{(u)_l}|\mathbf{r}_u)\) represent the output of the local encoder \(q_{\varphi_u}\). Therefore, we can express \(\mathbf{z}^{(u)}\) as:
\begin{equation}
    \mathbf{z}^{(u)} = w_{u1} \cdot \mathbf{z}^{(u)}_g + w_{u2} \cdot \mathbf{z}^{(u)}_l
\end{equation}
where the weights \(w_{u1}\) and \(w_{u2}\) are generated by the gating network \(h_{\psi_u}\) based on the user interaction data \(\mathbf{r}_u\) on each client. Therefore, by citing Lemma 1, we have:
\begin{equation}
    \begin{split}
        \bar{\mu}(\mathbf{r}_u) = \omega_{u1} \cdot \mu_{\varphi}(\mathbf{r}_{u}) + \omega_{u2} \cdot \mu_{\varphi_u}(\mathbf{r}_{u}), \\
        \bar{\sigma}^2(\mathbf{r}_u) = \omega_{u1}^2 \cdot \sigma^2_{\varphi}(\mathbf{r}_{u}) + \omega_{u2}^2 \cdot \sigma^2_{\varphi_u}(\mathbf{r}_{u}).
    \end{split}
\end{equation}
Thus, we have derived Eq. (4) of the main paper.

\subsection{Algorithm Optmization}
In Algorithm 1, $\varphi$ and $\theta$ are updated using the accumulated gradients \(\nabla_{\varphi}^{(u)} q\) and \(\nabla^{(u)}_{\theta} p\) uploaded by each client through the gradient descent algorithm. In this section, we will briefly discuss the rationale behind this update method.

From lines 3 and 4 of the $\operatorname{ClientUpdate}$ function in Alg. 1, it is evident that \(\varphi\) and \(\theta\) are updated locally on each client. Taking \(\theta\) as an example, let \(\theta^{t}\) be the value of \(\theta\) after the \( t \)-th local iteration. The update rule can be written as: 
\begin{equation}
    \begin{aligned}
        \theta^{(t)} & =\theta^{(t-1)}-\eta \cdot \nabla_{\theta} f^{(t)} \\
        & =\theta^{(t-2)}-\eta \cdot \sum_{i=t-1}^{t} \nabla_{\theta} \mathcal{L}_{\beta} \\
        & =\cdots \\
        & =\theta^{(0)}-\eta \cdot \sum_{i=1}^{t} \nabla_{\theta} \mathcal{L}_{\beta},
    \end{aligned}
\end{equation}
where \(\theta^{(0)}\) is initialized by the client using the received \(\theta\) for the current communication round. Since the value of \(\eta\) remains constant during local updates, line 6 of the \(\operatorname{ClientUpdate}\) function in Alg. 1 can be expressed as: 
\begin{equation}
    \nabla_{\theta} = \sum_{i=1}^{t} \nabla_{\theta} \mathcal{L}_{\beta} = \frac{1}{\eta}(\theta^{(0)} - \theta^{(t)}).
\end{equation}
Therefore, the accumulated gradient \(\nabla_{\theta}\) can be used to update \(\theta\) in the Global Procedure. Similarly, the accumulated gradient \(\nabla_{\varphi}\) can be used to update \(\varphi\) in the Global Procedure.

\section{Convergence Analysis}
In this section, We draw on the convergence analysis from work \cite{tan2022fedproto,yi2024fedmoe} to discuss the convergence of FedDAE.
For notational simplicity, we use \(\mathbf{\Theta}_u\) to represent \(\{\varphi, \varphi_u, \psi_u, \theta\}\),
use \(\mathcal{L}_u^{(t)} = \mathcal{L}_{\beta} (\mathbf{r}_u; \mathbf{\Theta}_u^{(t)})\) for the $u$-th user at the $t$-th communication round, and let $g_{u}^{(t)} = \nabla \mathcal{L}_{\beta} (\mathcal{B}_u^{(t)}; \mathbf{\Theta}_u^{(t)})$, where \(\mathcal{B}_u^{(t)}\) is a batch of local data \(\mathbf{r}_u\) at the \(t\)-th communication round.
We denote \(e \in \{0, 1/2, 1, 2, \cdots, E\}\) as the local iteration, and \(tE+e\) is the \(e\)-th local update in the $(t+1)$-th communication round.  \(e=0\) denotes that the time step when the client receives the global shared parameters \(\varphi\) and \(\theta\). 

\begin{assumption}[Lipschitz Smoothness]
    \label{ass:lip}
    Gradients of the $u$-th user's local model are \(L_1\)-Lipschitz continuous, i.e.,
    \begin{equation}
        \begin{array}{cc}
            \|
                \nabla \mathcal{L}_u^{(t_1)} - 
                \nabla \mathcal{L}_u^{(t_2)}
            \|_2 \leq 
            L_1 \| \mathbf{\Theta}_u^{(t_1)} - \mathbf{\Theta}_u^{(t_2)} \|, \\
            \forall t_1, t_2 > 0, u \in \{1,2,\cdots,n\}.
        \end{array}
    \end{equation}
    The above formula can be further derived as the following quadratic bound:
    \begin{equation}    
    \resizebox{1\linewidth}{!}{$
        \begin{array}{cc}
            \mathcal{L}_u^{(t_1)} - \mathcal{L}_u^{(t_2)} \leq
            \langle \nabla \mathcal{L}_u^{(t_2)}, (\mathbf{\Theta}_u^{(t_1)} - \mathbf{\Theta}_u^{(t_2)}) \rangle +
            \frac{L_1}{2} \| \mathbf{\Theta}_u^{(t_1)} - \mathbf{\Theta}_u^{(t_2)} \|
            , \\
            \forall t_1, t_2 > 0, u \in \{1,2,\cdots,n\}.
        \end{array}
    $}
    \end{equation}
    
\end{assumption}

\begin{assumption}[Unbiased Gradient and Bounded Variance]
\label{ass:unbias}

The stochastic gradient \(g_{u}^{(t)}\) is unbiased, i.e.,
\begin{equation}
    \mathbb{E}_{\mathcal{B}_u^{(t)} \subseteq \mathbf{r}_u}[g_{u}^{(t)}] = \nabla \mathcal{L}_u^{(t)}, 
    \forall u \in \{1,2,\cdots,n\},
\end{equation}
and the variance is bounded by:
\begin{equation}
    \mathbb{E} [
    \| g_{u}^{(t)} - \nabla \mathcal{L}_u^{(t)} \|^2_2
    ] \leq \tau^2.
\end{equation}

\end{assumption}

\begin{assumption}[Bounded Parameter Variation]
\label{ass:bounded}
The parameter variation of the global components \(\upsilon=\{\varphi, \theta\}\) before and after aggregation is bounded as:
\begin{equation}
    \| \upsilon^{(t)} - \upsilon^{(t)}_u \|^2_2 \leq \delta^2,
\end{equation}
where \( \upsilon^{(t)}_u \) is updated on the \(u\)-th client at the \(t\)-th communication round.    
\end{assumption}

Based on the above assumptions, we have the following Lemmas according to the work \cite{tan2022fedproto,yi2024fedmoe}.

\begin{lemma}[Local Model Training]
    \label{lemma:local}
    When all the above assumptions hold, for an arbitrary client's model in the $(t+1)$-th communication round, we have:
    \begin{equation}
        \resizebox{0.99\linewidth}{!}{$
        \mathbb{E}[\mathcal{L}_u^{(t+1)}] \leq 
        \mathcal{L}_u^{(tE+0)} + (\frac{L_1 \eta^2}{2} - \eta) \sum_{e=0}^E \| \nabla \mathcal{L}_u^{(tE+e)} \|^2_2 +
        \frac{L_1 E \eta^2 \tau^2}{2}.
        $}
    \end{equation}
\end{lemma}

\begin{lemma}[Server Aggregation]
    \label{lemma:agg}
    When Assumption \ref{ass:unbias} and \ref{ass:bounded} hold, after the $(t+1)$-th communcation round, the loss of any client before and after aggregating the shared parameter $\upsilon$ at the server is bounded by:
    \begin{equation}
        \mathbb{E}[\mathcal{L}_u^{((t+1)E+0)}] \leq 
        \mathbb{E}[\mathcal{L}_u^{(tE+1)}] + \eta \delta^2.
    \end{equation}
\end{lemma}

For the detailed proof of these lemmas, please refer to the work \cite{tan2022fedproto,yi2024fedmoe}.
When FedDAE meets the above assumptions, we can proceed with the following discussions based on the lemmas \ref{lemma:local} and \ref{lemma:agg}.

\begin{discus}[One-round deviation]
    \label{dis:one}
    Based on Lemma \ref{lemma:local} and \ref{lemma:agg},
    for any client, after the stages of local training, server aggregation and receiving the new global shared parameters, we have:
    \begin{equation}
        \label{eq:one_round}
        \resizebox{0.99\linewidth}{!}{$
        \begin{split}
            & \mathbb{E}[\mathcal{L}_u^{((t+1)E+0)}] \leq \\
            &\mathcal{L}_u^{(tE+0)} + (\frac{L_1 \eta^2}{2} - \eta) \sum^E_{e=0} \| \nabla \mathcal{L}_u^{(tE+e)} \|^2_2 + 
        \frac{L_1 E \eta^2 \tau^2}{2} + \eta \delta^2.
        \end{split}
        $}
    \end{equation}
\end{discus}

\begin{proof}
    By substituting the right-hand side of the inequality in Lemma \ref{lemma:agg} with Lemma \ref{lemma:local}, we can directly obtain Eq. \ref{eq:one_round}.
\end{proof}

\begin{discus}[Non-convex Convergence rate of FedDAE]
    \label{dis:converge}
    Based on the Theorem \ref{dis:one}, for any client \(u\) and a constant \(\rho>0\), we have the following:
    \begin{equation}
        \resizebox{1\linewidth}{!}{$
        \begin{array}{ll}
            & \frac{1}{T} \sum_{t=0}^{T-1} \sum_{e=0}^{E-1} \| \nabla \mathcal{L}_u^{(tE+e)} \|^2_2 \leq \\
            & \frac{
            \frac{1}{T} \sum_{t=0}^{T-1} [
                \mathcal{L}_u^{(tE+0)} - \mathbb{E}[\mathcal{L}_u^{((t+1)E+0)}]
            ] + \frac{L_1 E \eta^2 \tau^2}{2} + \eta \tau^2
            }{\eta - \frac{L_1 \eta^2}{2}} 
            < \rho, \\
             & \operatorname{s.t. } \eta < \frac{2(\rho - \delta^2)}{L_1 (\rho + E \delta^2)}.
        \end{array}
        $}
    \end{equation}
\end{discus}
From Theorem \ref{dis:converge},  it can be seen that any client model of FedDAE can converge at a non-convex rate of \(O(\frac{1}{T})\).

\begin{proof}
First, by bringing term \(\sum^E_{e=0} \| \nabla \mathcal{L}_u^{(tE+e)} \|^2_2\) from Eq. \ref{eq:one_round} to the left side of the inequality, we obtain:
\begin{equation}
    \resizebox{1\linewidth}{!}{$
        \begin{split}
            \sum_{e=0}^{E-1} \| \nabla \mathcal{L}_u^{(tE+e)} \|^2_2 \leq
            \frac{
                \mathcal{L}_u^{(tE+0)} - \mathbb{E}[\mathcal{L}_u^{((t+1)E+0)}
            ] + \frac{L_1 E \eta^2 \tau^2}{2} + \eta \tau^2
            }{\eta - \frac{L_1 \eta^2}{2}}.
        \end{split}
    $}
\end{equation}

Let $\mathcal{L}_u^{*}$ be the optimal objective of the $u$-th client, we have:
\begin{equation}
    \sum^{T-1}_{t=0} [\mathcal{L}_u^{(tE+0)} - \mathbb{E}[\mathcal{L}_u^{((t+1)E+0)}]] \leq \mathcal{L}_u^{(0)} - \mathcal{L}_u^{*}.
\end{equation}
Thus, we can get:
\begin{equation}
    \resizebox{1\linewidth}{!}{$
    \frac{1}{T} \sum_{t=0}^{T=1} \sum_{e=0}^{E-1} \| \nabla \mathcal{L}_u^{(tE+e)} \|^2_2 \leq
    \frac{
        \frac{1}{T} (\mathcal{L}_u^{(0)} - \mathcal{L}_u^{*})
        + \frac{L_1 E \eta^2 \tau^2}{2} + \eta \tau^2
    }{\eta - \frac{L_1 \eta^2}{2}}.
    $}
\end{equation}
Assume that the above equation converges to a constant $\rho>0$, the inequality becomes:
\begin{equation}
    \resizebox{1\linewidth}{!}{$
    \frac{1}{T} \sum_{t=0}^{T=1} \sum_{e=0}^{E-1} \| \nabla \mathcal{L}_u^{(tE+e)} \|^2_2 \leq
    \frac{
        \frac{1}{T} (\mathcal{L}_u^{(0)} - \mathcal{L}_u^{*})
        + \frac{L_1 E \eta^2 \tau^2}{2} + \eta \tau^2
    }{\eta - \frac{L_1 \eta^2}{2}} < \rho.
    $}
\end{equation}
Then,
\begin{equation}
    T > 
    \frac{\mathcal{L}_u^{(0)} - \mathcal{L}_u^{*}}{
        \rho (\eta - \frac{L_1 \eta^2}{2}) - \frac{L_1 E \eta^2 \tau^2}{2} - \eta \tau^2
    }.
\end{equation}
Since $T > 0$ and $\mathcal{L}_u^{(0)} - \mathcal{L}_u^{*} > 0$, we can know that:
\begin{equation}
    \rho (\eta - \frac{L_1 \eta^2}{2}) - \frac{L_1 E \eta^2 \tau^2}{2} - \eta \tau^2 > 0.
\end{equation}
By solving the above inequality, we finally get:
\begin{equation}
    \eta < \frac{2(\rho - \delta^2)}{L_1 (\rho + E \delta^2)}.
\end{equation}
Since $\rho$, $L_1$, $\tau^2$ and $\delta^2$ are all constants that greater than 0, $\eta$ must have solutions.
Thus, when the learning rate $\eta$ satisfies the above condition, and client's local model can converge.
We can easily derive that the non-convex convergence rate of FedDAE is \(O(\frac{1}{T})\).
\end{proof}

\section{More Experiment Details}


\subsection{Experimental Setting}
Table \ref{table:datasets} provides the statistical details of the datasets employed in this study includes the following:
\#Ratings represents the number of observed ratings.
\#Users denotes the number of users.
\#Items indicates the number of items.
Sparsity is the percentage of \#Ratings out of the total possible ratings.

All the datasets used in the experiments are publicly available:
MovieLens-100K (ML-100K)\footnotemark[1], MovieLens-1M (ML-1M)\footnotemark[1], Amazon-Instant-Video (Video)\footnotemark[2], and QB-article\footnotemark[3].
All methods were implemented using PyTorch \citep{paszke2019pytorch}, and experiments were conducted on a machine equipped with a 2.5GHz 14-Core Intel Core i9-12900H processor, a RTX 3070 Ti Laptop GPU, and 64GB of memory.

\footnotetext[1]{\url{https://grouplens.org/datasets/movielens/}}
\footnotetext[2]{\url{http://jmcauley.ucsd.edu/data/amazon/}}
\footnotetext[3]{\url{https://github.com/yuangh-x/2022-NIPS-Tenrec}}

\subsection{More Ablation Study}
By varying different components of the FedDAE model, we aim to investigate the impact of each component on the model's performance in this section:
(1) \textbf{FedDAE$_{global}$}: Aggregates the gradients of all components on the server for global aggregation;
(2) \textbf{FedDAE$_{local}$}: Keeps all component information local without sharing it with the server;

\begin{figure}[!htbp]
    \centering
    \includegraphics[width=0.9\linewidth]{figures/[movielens.ml-100k] global_local of FedMAE.pdf}
    \caption{
        Comparison of the performance of FedDAE with FedDAE$_{local}$ and FedDAE$_{global}$ on the ML-100K dataset. 
    }
    \label{fig:global_local}
\end{figure}

\paragraph{Ablation Study on Locality and Sharing.}
Fig. \ref{fig:global_local} shows the performance of FedDAE and its two variants, FedDAE$_{local}$ and FedDAE$_{global}$, on the ML-100K dataset in terms of HR@20 and NDCG@20 as a function of iterations. As illustrated in the figure, FedDAE achieves the best performance on both HR@20 and NDCG@20 metrics, indicating that its strategy of combining global and local encoders with a gating network is effective. FedDAE$_{local}$ performs poorly due to learning only from local data, while FedDAE$_{global}$, although improved, still suffers from instability and lower effectiveness due to the lack of personalized information for users.

\section{Limitation Discussion}

The parameters learned by the two encoders in FedDAE are related to the number of items \( m \) in the item set, which may require substantial storage space in practical applications. However, the generative capabilities of VAE might offer a potential solution, such as modeling the feature space of items to generate new ones. Additionally, the FedDAE architecture includes a global encoder, a local encoder, and a gating network, making the model relatively complex. As the number of clients increases, the overall time complexity and space complexity will also significantly increase.

Although FedDAE performs well on datasets with lower sparsity and moderate size (such as ML-100K and ML-1M) as shown in Figure 3, its performance on highly sparse datasets (such as Video and QB-article) is less satisfactory. Therefore, additional strategies may be required to handle such highly sparse datasets.


\bibliography{aaai25}